\documentclass[twocolumn]{aastex631}

\usepackage{amsmath}

\submitjournal{ApJ}

\begin{document}

\title{Latitude Quenching Nonlinearity in the Solar Dynamo}

\author[0000-0002-2728-4053]{Anthony R. Yeates}
\affiliation{Department of Mathematical Sciences, Durham University, Durham DH1 3LE, UK}

\author[0000-0002-1155-7141]{Luca Bertello}
\affiliation{National Solar Observatory, Boulder, CO 80303, USA}

\author[0000-0001-9746-9937]{Alexander A. Pevtsov}
\affiliation{National Solar Observatory, Boulder, CO 80303, USA}

\author[0000-0003-0489-0920]{Alexei A. Pevtsov}
\affiliation{National Solar Observatory, Boulder, CO 80303, USA}

\correspondingauthor{Anthony R. Yeates}
\email{anthony.yeates@durham.ac.uk}

\begin{abstract}

We compare two candidate nonlinearities for regulating the solar cycle within the Babcock-Leighton paradigm: tilt quenching (whereby the tilt of active regions is reduced in stronger cycles) and latitude quenching (whereby flux emerges at higher latitudes in stronger solar cycles). Digitized historical observations are used to build a database of individual magnetic plage regions from 1923 to 1985. The regions are selected by thresholding in \ion{Ca}{2} K synoptic maps, with polarities constrained using Mount Wilson Observatory sunspot measurements. The resulting data show weak evidence for tilt quenching, but much stronger evidence for latitude-quenching. Further, we use proxy observations of the polar field from faculae to construct a best-fit surface flux transport model driven by our database of emerging regions. A better fit is obtained when the sunspot measurements are used, compared to a reference model where all polarities are filled using Hale's Law. The optimization suggests clearly that the ``dynamo effectivity range'' of the Sun during this period should be less than $10^\circ$; this is also consistent with latitude quenching being dominant over tilt quenching.
\end{abstract}


%%%%%%%%%%%%%%%%%%%%%%%%%%%%%%%%%%%%%%%%%%%%%%%%%%%%%%%%%%
\section{Introduction} \label{sec:intro}

The number of visible sunspots is well-known to rise and fall quasi-regularly every 10 to 11 years, tracking the underlying magnetic cycle that pervades all aspects of solar activity \citep{hathaway2015, clette2023}. Yet we still lack a definitive explanation for what causes fluctuations in amplitude from one cycle to the next, or for the physical origin of the nonlinearities that prevent runaway exponential growth or decay \citep{petrovay2020a}.

In the currently-favored Babcock-Leighton paradigm \citep{babcock1961, leighton1969, charbonneau2007, choudhuri2023, cameron2023, karak2024}, the solar cycle represents an alternation between poloidal and toroidal components of the magnetic field. The ``winding up'' of the poloidal field into toroidal field by differential rotation is understood to be an essentially linear process: a high polar field will lead to more toroidal field in the following cycle, producing more sunspots. This is supported by good statistical correlations between the polar field at cycle minimum and the amplitude of the following cycle \citep{munoz-jaramillo2013}. It follows that the required nonlinearity preventing runaway cycle growth must lie in the toroidal-to-poloidal phase of the cycle. However, the production of net poloidal field by decaying active regions is also thought to be a primarily linear process. This is supported by the success of classical surface flux transport models in matching the observed polar field evolution  \citep[see reviews by][]{sheeley2005, yeates2023}. By elimination, we therefore expect the dominant nonlinearity to lie in the formation of active regions from the underlying toroidal field \citep[e.g.,][]{charbonneau2023}.

In this paper, we consider two candidate mechanisms visible directly at the solar surface: tilt quenching and latitude quenching \citep{petrovay2020a}. It should be stressed that these are by no means the only possible nonlinearities acting in the dynamo loop, even within the Babcock-Leighton paradigm. But other possibilities are beyond the scope of this paper and are mostly still at the stage of theoretical investigation, rather than observational testing. For a review, see \citet{karak2023} or \citet{cameron2023}.

Tilt quenching is the notion that active regions in higher-amplitude cycles tend to be less tilted with respect to the equator. This could therefore saturate the polar field production as cycle amplitude increases. Tilt quenching is expected theoretically from the picture of rising flux tubes in the convection zone \citep{dsilva1993, fan2009, karak2017}; a similar effect could also result from flux-dependent inflows towards active regions on the surface \citep{cameron2012}. Observational evidence has been harder to find and more controversial \citep{dasiespuig2010, dasiespuig2013, ivanov2012,  mcclintock2013, wang2015, tlatova2018, jha2020}, not least because there is so much scatter of active region tilts, and they evolve over the lifetime of an active region. However, the recent detailed analysis of \citet{jiao2021} for white-light sunspots did find a weak but significant negative correlation between the slope of Joy's Law (tilt versus latitude) and solar cycle amplitude.

Latitude quenching is based on the longstanding observation that, on average, active regions emerge at higher latitudes in stronger cycles \citep{waldmeier1939, li2003, solanki2008, tlatov2010, jiang2011}. This leads to saturation of the polar field because less of the leading polarity flux can escape across the equator, leaving less net polar field. The potential for this effect to regulate the solar cycle was apparently not noted in the literature until \citet{jiang2020}. Subsequent theoretical work by \citet{talafha2022} suggests that both of these nonlinearities could play a role, with their relative contribution depending on the nature of the active region decay and flux transport process.

One difficulty with choosing between latitude and tilt quenching (or indeed other nonlinearities) is that they are strongly affected by stochastic fluctuations between individual active regions. Indeed, recent work suggests that a small number of so-called ``rogue'' active regions (near the equator, lots of flux and/or highly tilted) can have a disproportionate effect on the polar field \citep{cameron2014, nagy2017}. So it does not suffice to consider only typical, or random, active region properties. Rather, account must be taken of all the particular active regions within each observed solar cycle. This cannot yet be done conclusively with magnetogram observations because these systematically date back only to Solar Cycle 21 (SC21). Previous historical work \citep[e.g.][]{jiang2011} has therefore relied on white-light sunspots, for which records date back to at least 1607 \citep{hayakawa2024}, although tilt angles are available only since about 1915 (SC15).  However, it is known sunspot tilt angles do not precisely correspond with those of magnetic regions \citep{wang2015}, which are ultimately more fundamental for the polar field production.

In this paper, we use digitized historical observations from Mount Wilson Observatory to reconstruct a database of magnetic regions from 1923 to 1985. The groundwork has been laid in a series of papers \citep{bertello2020, pevtsov2016, virtanen2019, virtanen2022}. Each region in our database covers the full spatial extent of strong flux, rather than being limited purely to white-light sunspots as in many of the previous studies. This is potentially important for accurately assessing their individual tilts or axial dipole strengths. The construction of our database is described in Section \ref{sec:extraction}, along with its calibration/testing against observed magnetograms during the overlap period in SC21.

Two complementary approaches are then used to consider tilt and latitude quenching. In Section \ref{sec:quenching}, direct evidence for each effect is sought through appropriate correlations -- this connects with most of the observational literature. However, we also leverage recent theoretical work by \citet{talafha2022} to determine indirectly the relative importance of tilt versus latitude quenching. Specifically, we fit a surface flux transport model, driven by our magnetic regions, to proxy observations of the Sun's polar fluxes (Section \ref{sec:sft}). The resulting best-fit model parameters allow us, in Section \ref{sec:effectivity}, to determine whether tilt or latitude quenching is dominating. Section \ref{sec:conc} concludes.

%%%%%%%%%%%%%%%%%%%%%%%%%%%%%%%%%%%%%%%%%%%%%%%%%%%%%%%%%%
\section{Extraction of magnetic regions} \label{sec:extraction}

We follow the approach pioneered by \citet{virtanen2019}, with the following steps:
\begin{enumerate}
    \item Locations and shapes of individual active regions are derived by thresholding in \ion{Ca}{2} K synoptic maps (detailed in Section \ref{sec:plage}).
    \item Pixel polarities within these regions are informed, where possible, by sunspot polarity measurements (Section \ref{sec:pol}). The remainder are filled using Hale's polarity law. 
    \item Pixels within these regions are assumed to carry equal magnetic flux, with any positive/negative imbalance removed for every region (Section \ref{sec:flux}).
\end{enumerate}
The general idea that \ion{Ca}{2} K intensity correlates with magnetic flux is long-established \citep[e.g.,][]{babcock1955, harvey1999, mordvinov2020, chatzistergos2022}. In principle, using \ion{Ca}{2} K would allow us to extend a database of individual active regions back to the 1890s \citep{chatzistergos2020}, going beyond simply looking at white-light sunspots. \citet{virtanen2019} were the first to use the (newly-digitized) sunspot polarity measurements to model individual regions, and published a proof-of-concept flux transport model \citep{virtanen2022} driven by these regions over the 20th Century. In this paper, we refine their analysis and expand it to consider uncertainties, and more thorough parameter optimization of the flux transport model (Section \ref{sec:sft}). The application in this paper is to tilt and latitude quenching, but the database and model will have wider applicability.

The code used to generate our historical database is available open source \citep{code_sft-historical}.

%-----------------------------------------------------
\subsection{Plage extraction} \label{sec:plage}

\begin{figure*}
    \centering
    \includegraphics[width=\textwidth]{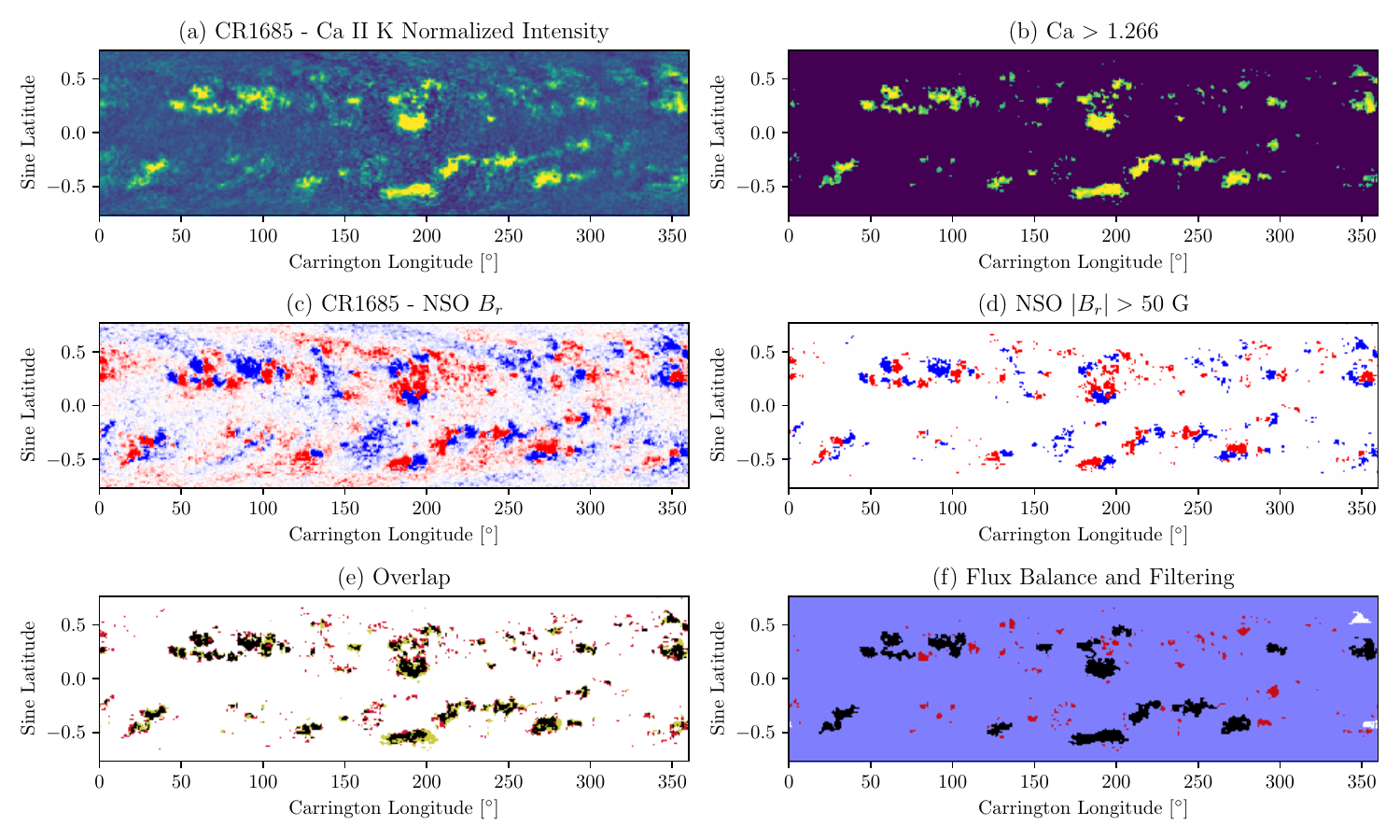}
    \caption{Comparison of synoptic maps for CR1685 (1979 August to September). Panel (a) shows the \ion{Ca}{2} K normalized intensity map while (b) shows only pixels exceeding $1.266$. Panel (c) shows $B_r$ from the NSO line-of-sight magnetogram, while (d) shows only pixels exceeding $\pm 50\,\mathrm{G}$. In (a,b), blue/yellow means low/high intensity, saturated at 0.8 and 1.6. In (c,d) red/blue means positive/negative, saturated at $\pm 50\,\mathrm{G}$. Panel (e) shows the overlap between these selected regions, where black pixels are selected in both maps, red only in the magnetogram, and yellow only in \ion{Ca}{2} K. Panel (f) indicates plages smaller than 50 pixels in red, larger flux-balanced plages in black ($\delta_{\rm unb}<0.5$), and larger unbalanced plages in white.}
    \label{fig:cak_cr1685}
\end{figure*}

Our starting point is the sequence of \ion{Ca}{2} K synoptic maps from \citet{bertello2020}, produced by digitizing, rescaling and combining daily spectroheliograms from Mount Wilson Observatory. To extract individual plage regions, we set a threshold of $1.266$ in the normalized \ion{Ca}{2} K intensity. As an example, the intensity map for Carrington rotation CR1685 (1979 August to September) is shown in Figure \ref{fig:cak_cr1685}(a), and after thresholding in Figure \ref{fig:cak_cr1685}(b).

The threshold has been determined by cross-comparison with synoptic magnetograms from US National Solar Observatory, Kitt Peak, during the overlap period covering CR1626 to CR1763 (1975 March to 1985 July), omitting rotations with incomplete maps. The magnetogram for CR1685 is shown in Figure \ref{fig:cak_cr1685}(c). Following \citet{virtanen2019}, we choose the threshold in normalized \ion{Ca}{2} K intensity that best corresponds to $|B_r|=50\,\mathrm{G}$ in the magnetograms. Specifically, we require the average percentage of selected pixels (over all rotations in the overlap period) to be equal to the average percentage of pixels exceeding $50\,\mathrm{G}$ in our set of NSO synoptic magnetograms, which is $3.77\%$. This gives $1.266$. Figure \ref{fig:cak_cr1685}(d) shows the pixels exceeding the magnetogram threshold, and these are seen to overlap quite well with the selected plages in Figure \ref{fig:cak_cr1685}(b). The overlap is illustrated in Figure \ref{fig:cak_cr1685}(e), and there is generally better agreement for larger/stronger magnetic regions.

Next, we cluster connected \ion{Ca}{2} K pixels into discrete plage regions. For our further analysis, the corresponding set of magnetogram pixels must be flux balanced. To assess the flux balance of each plage during the magnetogram overlap period, we calculate
\begin{equation} \label{eq:fluxbal}
    \delta_{\rm unb} = \left|\frac{\Phi_+ + \Phi_-}{\Phi_+ - \Phi_-}\right|,
\end{equation}
where $\Phi_{+}$ and $\Phi_{-}$ are the total positive and negative magnetic fluxes within the plage. Thus $\delta_{\rm unb}$ varies between $0$ for a perfectly balanced plage and $1$ for a perfectly unipolar plage. Many of the unbalanced/unipolar plages are small, and we opt (after trial and error) to filter out plages with fewer than 50 pixels -- shown red in Figure \ref{fig:cak_cr1685}(f). Across the overlap period, this leaves $85\%$ of plage pixels in ``balanced'' plage regions with $\delta_{\rm unb} < 0.5$, compared to $62\%$ before filtering. In Figure \ref{fig:cak_cr1685}(f), only two remaining plages are unbalanced (white). One is truly unipolar in the magnetogram while the other is really bipolar but is not selected accurately due to being near the threshold in \ion{Ca}{2} K. Since this filtering by size does not require magnetic data, we apply it also to the full historical dataset. However, it is not possible to remove the larger unbalanced plages without magnetic information, so this will give some uncertainty in our reconstruction.

Finally, it is necessary to remove plages where the \ion{Ca}{2} K synoptic maps are incomplete (Appendix \ref{app:gaps}). Rather than removing complete maps, we manually remove individual plages since the maps are often partially usable. In our final dataset for CR937 (1923 October) to CR1763 (1985 July), a total of 205 plages were removed for this reason, leaving 6910 remaining (1168 during the magnetogram overlap period of CR1626 to CR1763).

%-----------------------------------------------------
\subsection{Polarity assignment} \label{sec:pol}

Following \citet{virtanen2019}, we assign best-guess magnetic polarities to the extracted plage pixels using a combination of Hale's polarity rule and sunspot magnetic field measurements from Mount Wilson Observatory (MWO). This valuable archive of magnetic measurements stretches back to the year 1917, and has recently been digitized from the original sunspot drawings \citep{pevtsov2019, sunspotdata}.

\begin{figure*}
    \centering
    \includegraphics[width=\textwidth]{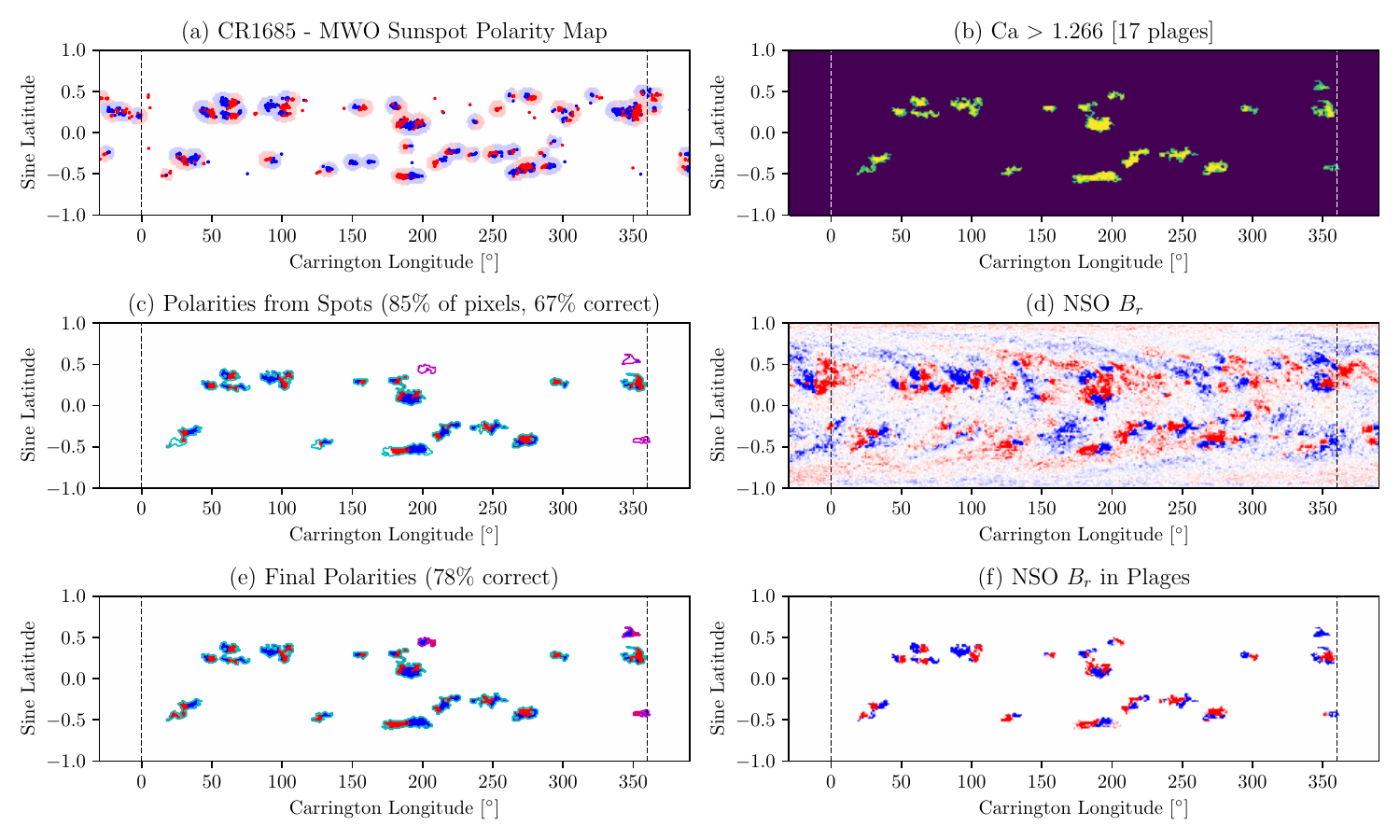}
    \caption{Assignment of polarities to plage pixels, for CR1685 (1979 August to September). Panel (a) shows the polarity map constructed from the sunspot measurements (see text), with red/blue meaning positive/negative. Panel (b) shows the plages identified from \ion{Ca}{2} K as in Figure \ref{fig:cak_cr1685}, while (c) shows these plage pixels with polarities assigned from (a). Panel (d) shows the corresponding NSO line-of-sight magnetogram (saturated at $\pm50\,\mathrm{G}$). Panel (e) shows the final reconstructed polarity map after filling unassigned polarities as described in the text, while (f) shows the ground-truth with plage pixels filled from the magnetogram (d). In (c,e), plages with complete/incomplete polarity information are outlined cyan/magenta.}
    \label{fig:polarities-nso}
\end{figure*}

We first correct a remaining longitude offset in the MWO sunspot positions (details in Appendix \ref{app:offsets}). These positions are then used to populate each pixel in an observed synoptic polarity map. Nearest-neighbor interpolation is used to label pixels as positive or negative according to the mean polarity of their $k$ nearest sunspots within maximum heliographic distance $\alpha$, similar to \citet{virtanen2019}. Taking $k=1$ would just assign the polarity of the nearest sunspot, but $k>1$ reduces fluctuations caused by errors in individual sunspot measurements. We find $\alpha=6^\circ$ and $k=5$ to maximize the number of correct pixel polarities in the magnetogram overlap period, so these values are adopted for the full dataset.

A polarity map is shown in Figure \ref{fig:polarities-nso}(a) for the example rotation CR1685, where dark pixels show the locations of actual sunspot measurements and lighter pixels lie within $6^\circ$ of a measurement. The plages for this rotation are shown in Figure \ref{fig:polarities-nso}(b), while Figure \ref{fig:polarities-nso}(c) shows the polarities that would be assigned to these plage pixels by the observed polarity map in Figure \ref{fig:polarities-nso}(a). For this rotation, $85\%$ of plage pixels are assigned a polarity by this technique. Since CR1685 is within the magnetogram overlap period, we can compare the reconstructed polarities to the observed synoptic magnetogram shown in Figure \ref{fig:polarities-nso}(d). Filling the plage pixels with the real polarities from the magnetogram gives the map in Figure \ref{fig:polarities-nso}(f); comparison of Figure \ref{fig:polarities-nso}(c) with this ground truth finds that $67\%$ of all plage pixels are assigned the correct polarity, with $18\%$ assigned the wrong polarity and the remaining $15\%$ unassigned.

To fill the unassigned polarities, our approach differs depending on whether the plage has sufficient information. If the plage has ``complete'' polarity information from the sunspot polarity map, meaning (i) at least one pixel of each polarity, and (ii) polarities for more than $50\%$ of pixels, then we fill in the remaining pixels using nearest-neighbor interpolation within the plage. These regions are outlined cyan in Figure \ref{fig:polarities-nso}(c). But if a region is incomplete, we throw away any polarity information from the sunspot map and populate the polarities using Hale's Law. Specifically, we first populate the east- and west- most pixels with opposite polarities, then fill in the remainder with nearest-neighbor interpolation in longitude. These regions are outlined magenta in Figure \ref{fig:polarities-nso}(c). To determine appropriate leading and following polarities, we separate neighboring solar cycles by interpolating the dataset of \citet{leussu2017}, who made a careful assignment of cycle number to individual sunspot groups.

\begin{figure}
    \centering
    \includegraphics[width=\columnwidth]{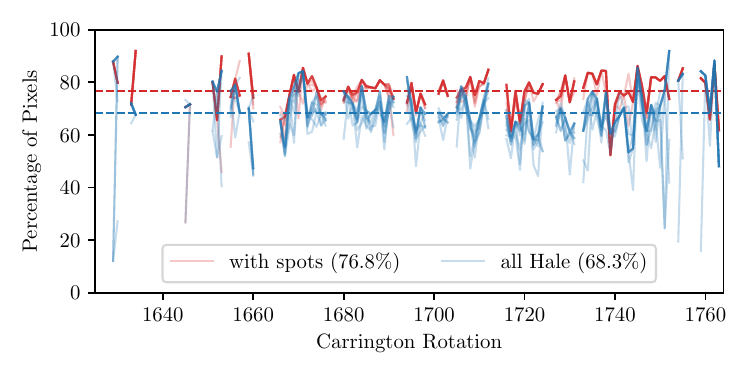}
    \caption{Percentage of plage pixels with correct polarities over the magnetogram overlap period (omitting rotations with incomplete data). Curves show the percentage of correctly assigned pixel polarities when sunspot measurements are used for polarity assignment (red) or not used (blue). Ten realizations are overlayed in each case (see text), with dashed lines showing overall means as indicated in the legend.}
    \label{fig:polarity-stats-nso}
\end{figure}

The results of polarity filling for CR1685 are illustrated in Figure \ref{fig:polarities-nso}(e), which increases the percentage of pixels having correctly reconstructed polarity information to $77\%$. As shown in Figure \ref{fig:polarity-stats-nso}, this success rate is typical for the magnetogram overlap period. Interestingly, Figure \ref{fig:polarity-stats-nso} also shows that there is only an $8.5\%$ reduction in accuracy if the sunspot measurements are discarded altogether and all plage pixels are filled using Hale's Law. However, there is a more significant effect on the axial dipole moment and polar field, as we will see later.

In fact, when the plage pixels are filled with the real magnetogram polarities, as in Figure \ref{fig:polarities-nso}(f), only $94\%$ of plages over the magnetogram overlap period obey Hale's Law (when comparing their polarity centroids). Accordingly, in our later calculations we randomly flip the polarity of $6\%$ of ``incomplete'' plages, and consider an ensemble of different realizations of this flipping. Ten such realizations are shown for each case in Figure \ref{fig:polarity-stats-nso}. It is evident from the figure that this random polarity flipping can lead to substantial fluctuations in the polarities in any given Carrington rotation -- for example around CR1760. This uncertainty is much reduced by using the sunspot measurements.

%-----------------------------------------------------
\subsection{Flux assignment} \label{sec:flux}

\begin{figure}
    \centering
    \includegraphics[width=\columnwidth]{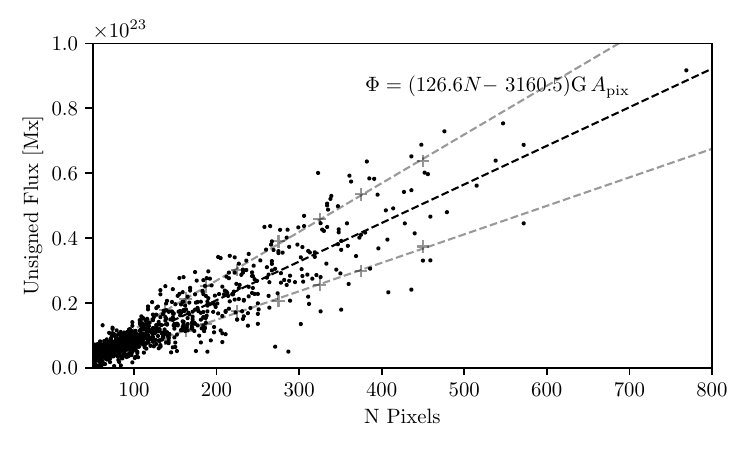}
    \caption{Scatterplot of (unsigned) observed magnetic flux, $\Phi$, in each plage against its size, for the magnetogram overlap period. The black dashed line shows the indicated linear fit, while the grey dashed lines show a linear fit to binned standard deviations ($+$ symbols). Here $A_{\rm pix}=9.39\times 10^{17}\,\mathrm{cm}^2$ is the (uniform) pixel area.}
    \label{fig:fluxfit-nso}
\end{figure}

Since the absolute magnetic field strengths in the sunspot dataset are less reliable than the polarities \citep{pevtsov2019}, we follow \citet{virtanen2019} and set each plage pixel to an equal field strength. This is calibrated using the magnetogram overlap period. Figure \ref{fig:fluxfit-nso} shows unsigned magnetic flux, $\Phi$, against area for each identified plage region between CR1626 and CR1763. In our reference dataset, we use the black-dashed linear fit to assign the field strength per pixel. However, there is a non-trivial spread of plage fluxes around this best fit, giving rise to a known uncertainty in our flux assignments. The effect of this uncertainty on our results will be evaluated by considering an ensemble of realizations with different randomly chosen fluxes taken from this distribution. Specifically, for a given plage size (number of pixels), the flux will be chosen from a normal distribution with mean following the black dashed line and standard deviation the grey dashed lines in Figure \ref{fig:fluxfit-nso}. After assigning the flux, each individual plage region is corrected for flux balance ($\Phi_+=-\Phi_-$), using a multiplicative correction to preserve pixel polarities. We denote the unsigned flux of a plage as $\Phi = 2\Phi_+ = -2\Phi_-$.

\begin{figure}
    \centering
    \includegraphics[width=\columnwidth]{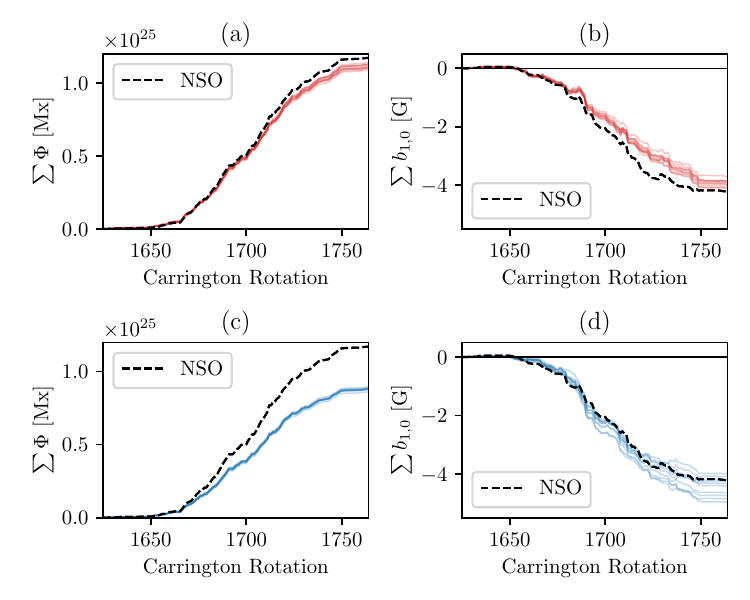}
    \caption{Cumulative plage fluxes, $\Phi$ (a,c) and axial dipole strengths, $b_{1,0}$ (b,d), during the magnetogram overlap period, compared to  magnetogram measurements (black dashed lines). Values are taken at emergence time only, neglecting any subsequent decay or surface flux transport. Only the 758 plages with (magnetogram) $\delta_{\rm unb} < 0.5$ are included. Top row (red curves) use sunspot data for polarity assignment, while bottom row (blue curves) do not. Faint lines show 10 realizations for each, differing in both the randomly chosen fluxes and which $6\%$ of incomplete plages have flipped polarity.
    All incomplete plages have $\Phi$ and $b_{1,0}$ reduced by a factor 1.3.}
    \label{fig:dipoles-nso}
\end{figure}

The cumulative plage fluxes for the magnetogram overlap period are shown by the red curves in Figure \ref{fig:dipoles-nso}(a). All ensemble members remain near the black dashed line, which shows the cumulative fluxes when the same plage pixels are filled from observed magnetograms. The slight underestimate arises because the ``incomplete'' plage fluxes have been reduced by a factor 1.3. This is clearer for the blue curves in Figure \ref{fig:dipoles-nso}(c), where no sunspot data have been used and all plages treated as ``incomplete''. The reason for reducing the incomplete fluxes is to avoid over-estimating the cumulative axial dipole strength, plotted in Figures \ref{fig:dipoles-nso}(b,d).

The axial dipole strength is a more critical quantity for the Babcock-Leighton dynamo than the region flux. For an individual plage, it is computed as
\begin{equation} \label{eq:dipole}
    b_{1,0} = \frac{3}{4\pi}\int_0^{2\pi}\int_0^{\pi}B_r(\theta,\phi)\cos\theta\sin\theta\,\mathrm{d}\theta\,\mathrm{d}\phi,
\end{equation}
by setting $B_r(\theta,\phi)=0$ outside the individual plage. It is more sensitive than $\Phi$ because it also depends on the latitudinal distribution of pixel polarities within the region. For the magnetogram overlap period, we find that the polarity-filling procedure based on sunspot measurements reproduces quite accurately the cumulative axial dipole strength over all plages (Figure \ref{fig:dipoles-nso}b). But if no sunspot measurements are used, our bipolar polarity-filling procedure for incomplete plages overestimates $b_{1,0}$ by a factor $1.3$ (averaged over the ensemble). The overestimate would be even worse without flipping $6\%$ of the polarities, but flipping more would lead to an unrealistic number of anti-Hale plages. The remaining overestimate arises from the multipolar nature of some of the plages, which we cannot predict without individual polarity observations. As a pragmatic solution, we therefore reduce $\Phi$ by 1.3 for all incomplete plages throughout our historical dataset. We verified that varying this parameter does not significantly change our results in Section \ref{sec:sft}.
 
%-----------------------------------------------------
\subsection{Full Dataset} \label{sec:full}

\begin{figure*}
    \centering
    \includegraphics[width=\textwidth]{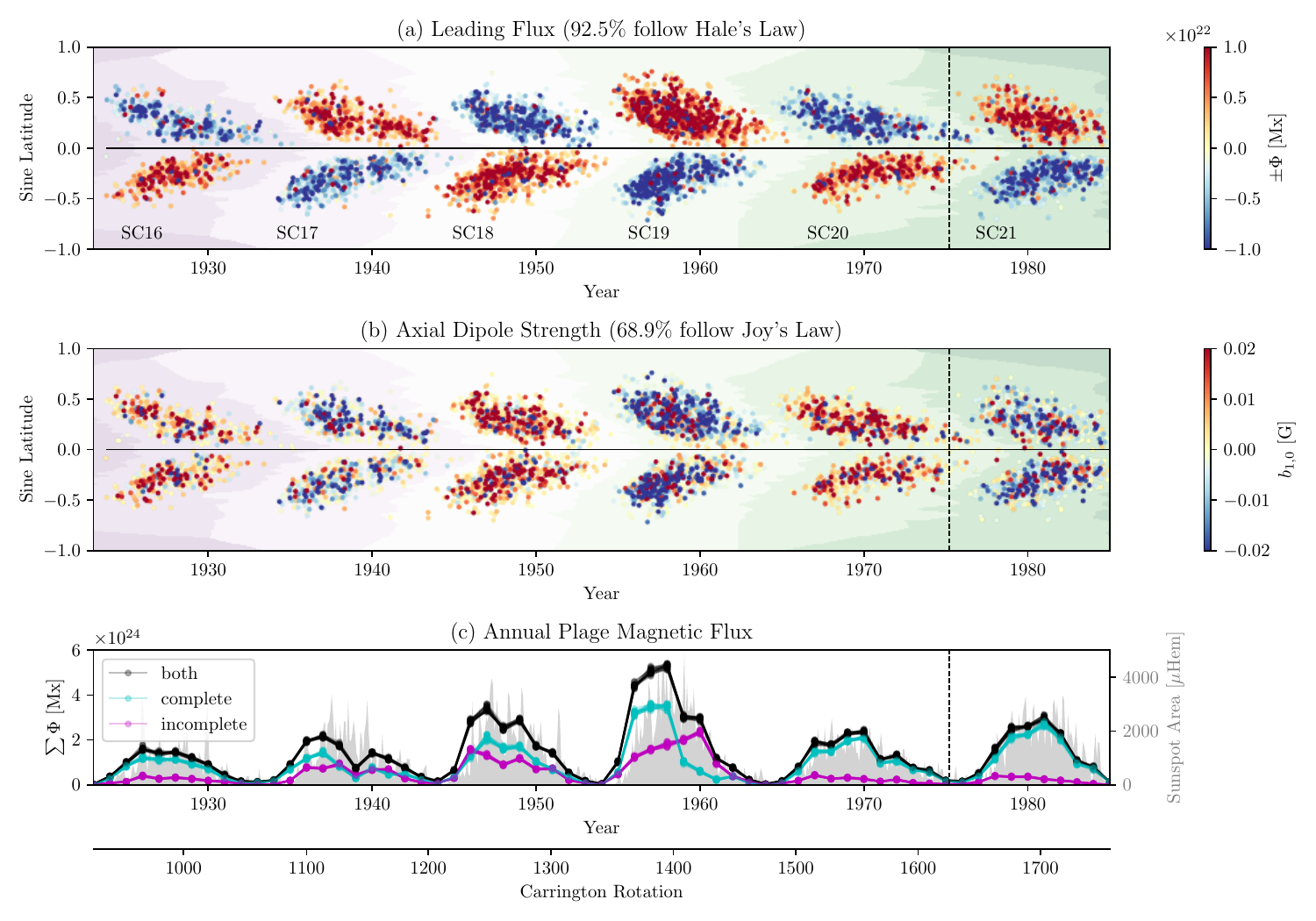}
    \caption{Reconstructed plages over the full period CR937 (1923 October) to CR1763 (1982 July). For a single realization, (a) shows plages in latitude-time colored by their flux (red/blue if the leading polarity is positive/negative), and (b) shows plages in latitude-time colored by their axial dipole strength. For all 20 realizations, (c) shows total plage flux in yearly bins, separately for ``complete'' plages with polarity information from sunspots (cyan) and for ``incomplete'' plages without (magenta). Combined fluxes for all plages are shown in dark gray, while light gray shading shows monthly-averaged sunspot areas from \citet{mandal2020}. Background purple/green shading in (a,b) shows the solar cycle numbers assigned by interpolating the data of \citet{leussu-data, leussu2017}. The start of the magnetogram overlap period in CR1626 (1975 March) is indicated with a dashed line.}
    \label{fig:regions-bfly}
\end{figure*}

Figure \ref{fig:regions-bfly} shows the full dataset of 6910 extracted magnetic regions, with (a,b) showing a single realization of the polarity and flux assignments. For this study, we generated an ensemble of 20 different realizations, differing in (i) which incomplete plages violate Hale's Law (the polarity flipping), and (ii) the individual plage fluxes for both complete and incomplete plages (accounting for the spread in Figure \ref{fig:fluxfit-nso}). But all show a similar overall pattern.

Figure \ref{fig:regions-bfly}(a) clearly shows Hale's Law, whereby the leading polarities are oppositely signed in each hemisphere and reverse from one cycle to the next. In this particular realization, $92.5\%$ of plages follow Hale's Law, while over the whole ensemble it is $92.3\pm0.2\%$. This is slightly below the observed $94\%$ in the magnetogram overlap period. Since we have fixed the rate for incomplete plages by the polarity flipping, this deficit over the longer dataset must arise from variation in complete plages.

Figure \ref{fig:regions-bfly}(b) shows Joy's Law, which here means that the sign of axial dipole strength, $b_{1,0}$, is on average the same for all regions in a cycle, and reverses from one cycle to the next. This is well-known to be only a statistical law, with broad scatter. In the particular realization shown, $68.9\%$ of plages have a sign of $b_{1,0}$ consistent with Joy's Law, while over the whole ensemble it is $68.7\pm0.2\%$. This compares with the observed $70\%$ during the magnetogram overlap period.  Interestingly, the sunspot polarity measurements are not necessary to recover Joy's Law: when all plages are treated as ``incomplete'' and the polarities filled using the Hale's Law procedure described in Section \ref{sec:pol} (including random polarity flipping), we find that $70.0\pm0.3\%$ of plages obey Joy's Law. This arises purely from the morphology of the plages, and the tendency for a systematic tilt with respect to the East-West line.

Figure \ref{fig:regions-bfly}(c) shows two things. Firstly, that the total amount of magnetic flux in the plages -- which is essentially proportional to the plage area -- tracks the total observed sunspot area quite well over time. This area is shown as a monthly average, computed from the calibrated daily sunspot areas of \citet{mandal2020}. Secondly, the separate curves for ``complete'' and ``incomplete'' plage fluxes illustrate how the coverage of the sunspot data varies over time. As we have seen, during the magnetogram overlap period (roughly SC21), the majority of flux comes from complete plages, and the same is true for SC20, as well as SC16. But for SC17-19, and SC19 in particular, a more substantial fraction of plages are incomplete. This figure shows all of the individual realizations, showing that the random fluxes have only a small effect overall. (The random polarity flipping in incomplete plages does not affect this plot.)

%%%%%%%%%%%%%%%%%%%%%%%%%%%%%%%%%%%%%%%%%%%%%%%%%%%%%%%%%%
\section{Evidence for quenching from magnetic regions} \label{sec:quenching}

Having assembled a dataset of magnetic regions, we first seek evidence of either latitude or tilt quenching purely from this dataset, without additional modeling assumptions. In particular, at this stage we do not consider subsequent evolution of their fluxes -- this will be addressed in Section \ref{sec:sft}.

%-----------------------------------------------------
\subsection{Latitude quenching} \label{sec:latq}

\begin{figure*}
    \centering
    \includegraphics[width=\textwidth]{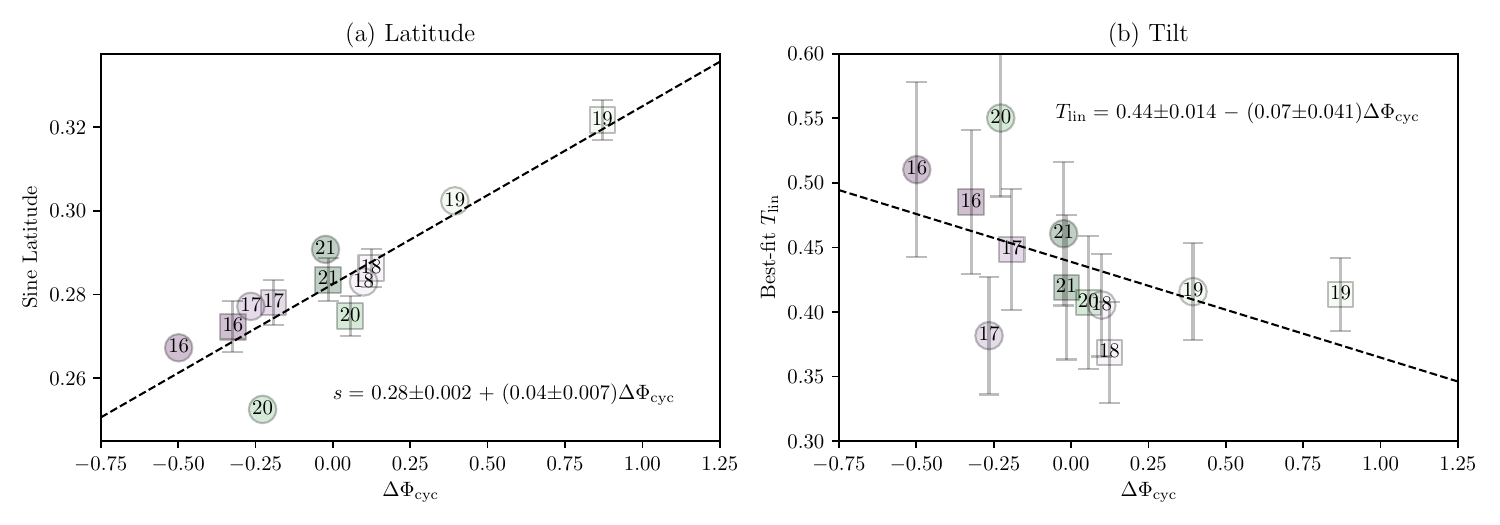}
    \caption{Scatterplots of (a) cycle-averaged plage sine-latitude centroids and (b) best-fit tilt coefficients, both against cycle strength -- here measured by the relative deviation of the total plage flux, $\Delta\Phi_{\rm cyc}$, from its mean over all cycles. Separate datapoints are shown for each hemisphere (North squares, South circles), with the vertical axis in (a) showing unsigned values so as to overlay the hemispheres. Dashed lines shows the indicated linear least-squares fits. Error bars in (a) show standard error of the mean, and in (b) show uncertainties in the fitted slopes, $T_{\rm lin}$. Pearson correlation coefficients are $r=0.88$ with $p=0.0002$ in (a), but only $r=-0.50$ with $p=0.1$ in (b).} 
    \label{fig:quenching-em}
\end{figure*}

We find a clear trend for plages in stronger solar cycles to be located at higher latitudes. This is illustrated by Figure \ref{fig:quenching-em}, where each datapoint corresponds to a single cycle and hemisphere (North/South). The vertical axis shows the flux-weighted centroid of sine latitude over all plages within a given cycle, while the horizontal axis shows the relative deviation of the total plage flux, $\Phi_{\rm cyc}$ from its all-cycle mean, $\overline{\Phi_{\rm cyc}}$. Specifically,
\begin{eqnarray}
    \Delta\Phi_{\rm cyc} = \frac{\Phi_{\rm cyc} - \overline{\Phi_{\rm cyc}}}{\overline{\Phi_{\rm cyc}}}.
\end{eqnarray}
This total plage flux is a direct proxy for cycle amplitude. We see a $25\%$ increase in average latitude from the lowest (SC20, South) to highest (SC19, North). The dashed line shows a linear fit (without omitting the outlier SC20 South). We note that, due to our flux assignment procedure, this result depends only on the plage locations and sizes, so is independent of the pixel polarity assignment and the associated uncertainty. The plage fluxes, $\Phi_{\rm cyc}$, do vary within the ensemble because of the random flux assignment, but the relative deviation (horizontal axis of Figure \ref{fig:quenching-em}) has standard deviation $<0.025$ for all cycles, so that error bars would fall within the plot symbols. The latitude (vertical axis) does not change between realizations.

In summary, there is clear evidence that active regions tend to emerge at higher latitudes in stronger cycles, in qualitative agreement with Figure 1(a) of \citet{jiang2020}.

%-----------------------------------------------------
\subsection{Tilt quenching} \label{sec:tiltq}

To assess whether there is a systematic reduction in active region tilts in stronger cycles, at a given latitude, we follow the ``unbinned fitting'' method of \citet{jiao2021}. To allow comparison with the previous literature, we define the tilt angle $\alpha\in[-90^\circ, 90^\circ]$ of a plage region using the polarity-agnostic definition of \citet{wang2015}. Namely,
\begin{eqnarray}
    \tan\alpha = \frac{\Delta\lambda}{\Delta\phi\cos\lambda_0},
\end{eqnarray}
where $\Delta\lambda$ is the latitude difference between the opposite polarity centroids, and $\Delta\phi$ is the longitude difference. Here $\lambda_0$ is the overall latitude centroid. The sign is chosen to be positive if the leading pole is equatorward of the following pole (in either hemisphere), negative otherwise.

Having computed the tilt angle $\alpha$ for each plage region, we separate regions in different hemispheres by their $\lambda_0$, and for each hemisphere use least-squares to find -- for each cycle -- the best-fit slope $T_{\rm lin}$ in the Joy's Law relation
\begin{equation}
    \alpha = T_{\rm lin}|\lambda_0|.
\end{equation}
We assume a linear functional form passing through the origin, in light of the analysis of \citet{jiao2021}. These best-fit slopes are plotted in Figure \ref{fig:quenching-em}(b) as a function of the relative cycle amplitude, $\Delta\Phi_{\rm cyc}$. As in previous studies, there is significant scatter in Joy's Law, leading to significant uncertainties in the best-fit slopes $T_{\rm lin}$, indicated by the error bars. Nevertheless, the linear fit (dashed line) does suggest a tilt-quenching trend of decreasing $\alpha$ in stronger cycles, although the negative correlation is weaker and much less statistically significant than the latitude-quenching effect. We also tried using weighted least-squares for fitting $T_{\rm lim}$, with uncertainties proportional to $\Phi^{-1/2}$, but this made the correlation even weaker.

%%%%%%%%%%%%%%%%%%%%%%%%%%%%%%%%%%%%%%%%%%%%%%%%%%%%%%%%%%
\section{Surface flux transport model} \label{sec:sft}

We have also verified that driving a surface flux transport (SFT) simulation with the extracted regions can give a consistent polar field evolution over the whole period. In this section, we describe the setup and calibration of this model. The implications for tilt and latitude quenching are discussed in Section \ref{sec:effectivity}.

That such a consistent model is possible was previously found by \citet{virtanen2022}, and used for example by \citet{lockwood2022}. Here we have carried out a more thorough optimization, using our refined dataset.

%-----------------------------------------------------
\subsection{Model equations} \label{sec:model}

We use the classical SFT model as reviewed in detail by \citet{yeates2023}. The radial magnetic flux density $B_r(\theta,\phi,t)$ on the solar surface $r=R_\odot$ obeys the advection-diffusion equation
\begin{equation}
    \frac{\partial B_r}{\partial t} + \nabla_h\cdot\left({\bf u} B_r\right) = \eta_0\nabla_h^2 B_r + S,
    \label{eqn:sft2d}
\end{equation}
where $\nabla_h$ denotes the surface gradient, ${\bf u}(\theta)$ the imposed large-scale (axisymmetric) surface flow, $\eta_0$ the supergranular diffusivity, $\nabla_h^2$ the Laplace-Beltrami operator, and $S(\theta,\phi,t)$ the source term. We emerge new regions instantaneously, so that
\begin{equation}
S(\theta,\phi,t) = \sum_{i=1}^{6910} B_r^{(i)}(\theta,\phi)\delta(t-t^{(i)}),
\end{equation}
where their individual magnetic fields $B_r^{(i)}$ and emergence times $t^{(i)}$ are taken from a given realization of our plage dataset. Unlike \citet{schrijver2002} or \citet{baumann2006}, who needed a radial decay term in Equation \eqref{eqn:sft2d} to reproduce regular polar field reversals, we do not find it necessary to include such a term. Nevertheless, we have verified that including an exponential decay term as an additional optimized parameter (not shown) does not change our conclusions.

Since the polar field proxy against which we will optimize is independent of longitude, it suffices to solve the longitude-average of (\ref{eqn:sft2d}), namely
\begin{eqnarray}
    &\displaystyle\frac{\partial\langle B_r\rangle}{\partial t} + \frac{1}{R_\odot\sin\theta}\frac{\partial}{\partial\theta}\Big(\sin\theta\,u_\theta\langle B_r\rangle\Big) = \nonumber\\
    & \displaystyle\frac{\eta_0}{R_\odot^2\sin\theta}\frac{\partial}{\partial\theta}\left(\sin\theta\frac{\partial\langle B_r\rangle}{\partial\theta}\right) + \langle S\rangle,
    \label{eqn:sft1d}
\end{eqnarray}
where $\langle B_r\rangle(\theta,t)$ is the longitude-average of $B_r(\theta,\phi,t)$. This is independent of the differential rotation $u_{\phi}(\theta)$, so the only relevant contribution to ${\bf u}$ is the meridional flow. We 
 assume a single peak in each hemisphere, with the two-parameter form
\begin{equation}
    u_{\theta}(\theta) = -R_\odot\Delta_u\cos\theta\sin^{p_0}\theta,
\end{equation}
from \citet{whitbread2018}, where
\begin{equation}
    \Delta_u = \frac{v_0(1+p_0)^{(1+p_0)/2}}{R_\odot p_0^{p_0/2}}
\end{equation}
is the flow divergence at the equator. This profile has two free parameters: $v_0$ is the maximum speed (at mid-latitudes), and $p_0$ controls the latitude of this maximum.

Equation (\ref{eqn:sft1d}) is solved on a uniform mesh of 180 cells in $\cos\theta$, using an explicit finite-volume method that conserves magnetic flux. Our open-source implementation is available at  \url{https://github.com/antyeates1983/sft-historical}. For the initial condition, at $t=t_0$ corresponding to 12:00 UT on 1923 October 31, we set
\begin{equation}
    \langle B_r\rangle (\theta, t_0) = \pm B_0\exp\left(-\frac{\mathrm{Rm}_0\sin^{1+p_0}\theta}{1+p_0}\right),
    \label{eq:steady}
\end{equation}
 which represents an approximate steady-state profile for the given model parameters, where $\mathrm{Rm}_0=R_\odot^2\Delta_u/\eta_0$ \citep[see][]{yeates2023}. Opposite signs are used in each hemisphere, appropriate for the start of SC16. The initial polar field strength $B_0\equiv \langle B_r\rangle(0,t_0)$ is assumed equal in both hemispheres and chosen by optimization since it is not directly observed. Each simulation runs continuously until 12:00 UT on 1985 July 31.

%-----------------------------------------------------
\subsection{Optimization method} \label{sec:optmethod}

For each plage realization, we have performed a brute-force exploration of parameter space with $10\,000$ runs, varying the four parameters $\eta_0\in[200, 1000]\,\mathrm{km}^2\,\mathrm{s}^{-1}$, $v_0\in[5, 30]\,\mathrm{m}\,\mathrm{s}^{-1}$, $p_0\in[1,10]$, and $B_0\in[-15, 0]\,\mathrm{G}$. Parameter sets were randomly selected by Latin hypercube sampling (using \url{https://pythonhosted.org/pyDOE/}). The whole exercise has been repeated twice: once where plage polarities use the full sunspot polarity measurements (where available), and once where all plages are treated as ``incomplete'' with polarities filled by Hale's Law.

\begin{figure*}
    \centering
    \includegraphics[width=\textwidth]{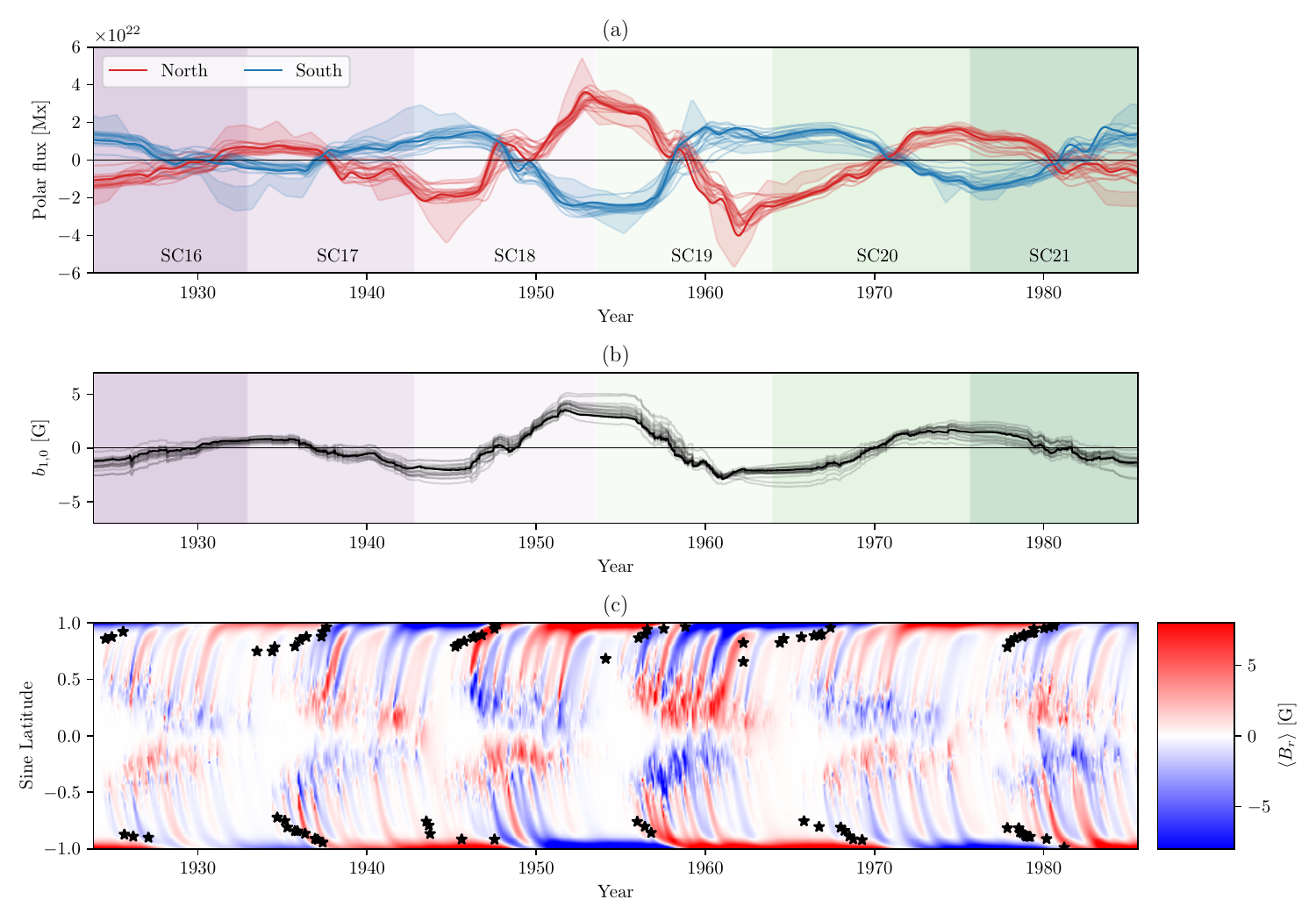}
    \caption{Best-fit SFT models for each of the 20 realizations in the plage ensemble using sunspot polarity measurements. Panel (a) shows North and South polar fluxes (poleward of $\pm 70^\circ$), (b) shows axial dipole strength, and (c) shows the longitude-averaged $B_r$ for the overall optimum model (corresponding to the darker curves in a, c). Red and blue shaded regions in (a) show the ground truth data from \citet{polarfielddata} with their provided uncertainties. Purple/green shading indicates solar cycle numbers. Black star symbols in (c) show the latitudes of polar crown filaments from \citet{xu2021} (see Section \ref{sec:pcf}).}
    \label{fig:opt-bfly-spots}
\end{figure*}

\begin{figure*}
    \centering
    \includegraphics[width=\textwidth]{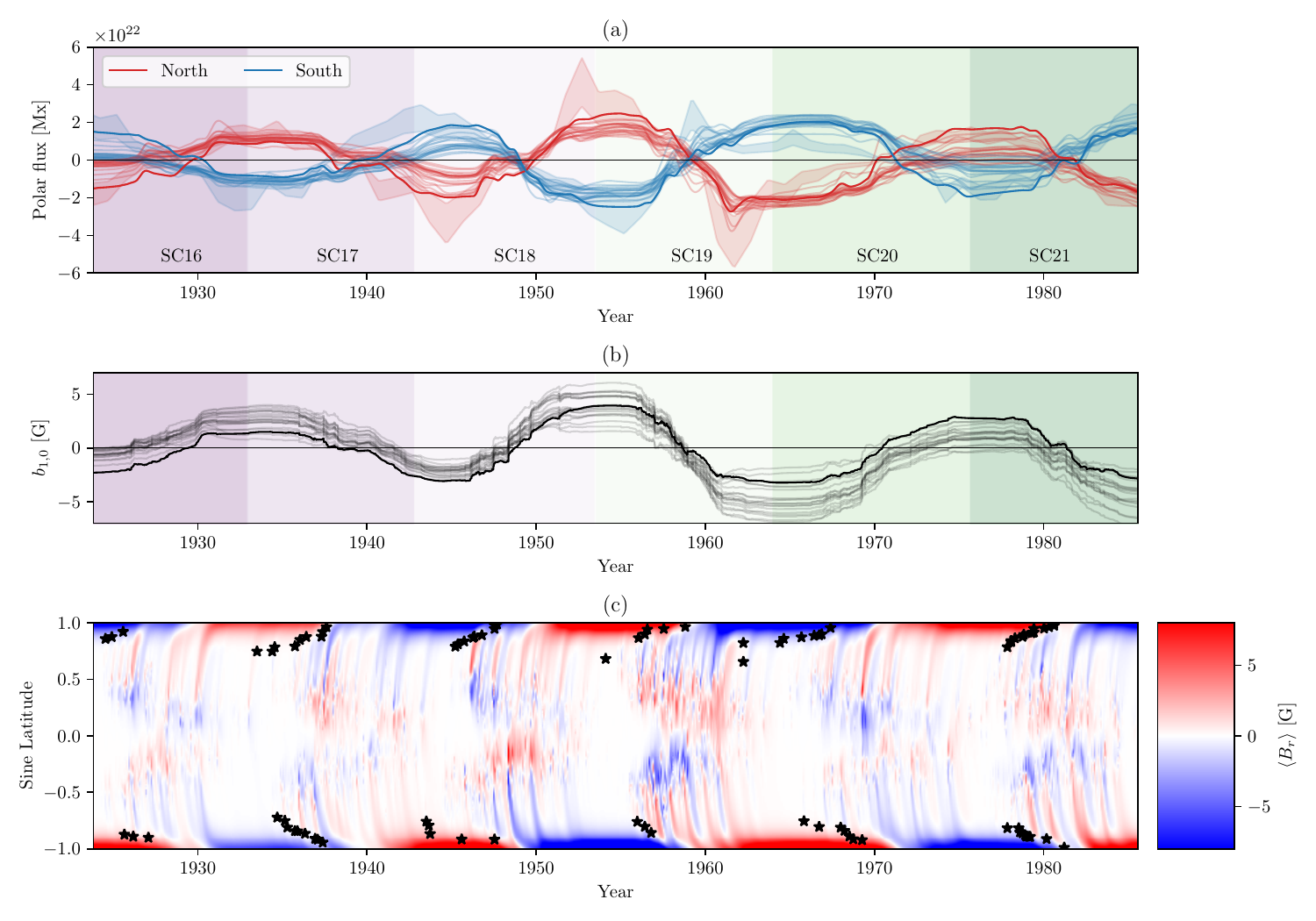}
    \caption{Same as Figure \ref{fig:opt-bfly-spots} but with all plage polarities filled using Hale's Law.}
    \label{fig:opt-bfly-hale}
\end{figure*}

Our assumed ground truth is the historical polar field reconstruction of \citet{munoz-jaramillo2012} \citep[available from][]{polarfielddata}, where calibrated counts of white-light polar faculae (bright points) from MWO have been used to infer North and South polar field strengths. The rationale is the original finding by \citet{sheeley1991} that faculae numbers are highly correlated with magnetogram polar field strengths. These proxies for the observed North and South polar fields are shown by the shaded regions in Figure \ref{fig:opt-bfly-spots}(a), repeated in Figure \ref{fig:opt-bfly-hale}(a), which incorporate the indicated uncertainties. The uncertainty in this proxy approach is also highlighted by comparing similar faculae counts from different observatories \citep[see][Fig. 11]{petrovay2020}.

To assess the fit of each SFT run, we compute the squared error of the North and South polar fluxes (above $\pm 70^\circ$ latitude) compared to the observational proxies,
\begin{eqnarray}
    \Delta\Phi^2_{\rm N}(t_k) &=& \left(\Phi_{\rm N}^{({\rm obs})}(t_k) - \Phi_{\rm N}^{({\rm sim})}(t_k)\right)^2,\\
    \Delta\Phi^2_{\rm S}(t_k) &=& \left(\Phi_{\rm S}^{({\rm obs})}(t_k) - \Phi_{\rm S}^{({\rm sim})}(t_k)\right)^2.
\end{eqnarray}
The polar flux error is then measured by the root-mean-square flux
\begin{equation}
    E_{\rm PF} = \sqrt{\frac{1}{n_t}\sum_{k}\Big[\Delta\Phi_{\rm N}^2(t_k) + \Delta\Phi_{\rm S}^2(t_k)\Big]},
\end{equation}
with equal weighting given to all of the equally-spaced sampling times, $t_k$.

%-----------------------------------------------------
\subsection{Optimization results} \label{sec:opt}

The individual curves in Figures \ref{fig:opt-bfly-spots}(a) and \ref{fig:opt-bfly-hale}(a) show the polar fields in best-fit SFT models using each of 20 different realizations of the plage regions (Section \ref{sec:extraction}). The difference between the two figures is whether or not sunspot polarity measurements were used during the plage polarity filling. The corresponding axial dipole strengths are shown in Figures \ref{fig:opt-bfly-spots}(b) and \ref{fig:opt-bfly-hale}(b), while time-latitude ``butterfly'' plots of $\langle B_r\rangle$ for the overall optimum run (lowest $E_{\rm PF})$ are shown in Figures \ref{fig:opt-bfly-spots}(c) and \ref{fig:opt-bfly-hale}(c).

As seen in Figure \ref{fig:opt-bfly-spots}(a), the optimum runs using sunspot measurements obtain quite a good match to the observed polar fields given the small number of free parameters. The worst agreement is in cycles SC16 (roughly 1925-1935) and SC21 (near the end of the simulation). In particular, in SC16, and despite the simulation starting with a relatively low polar field ($B_0$ parameter), there is a shortfall in polar field production delaying the reversal around 1928 and leading to an underestimated polar field during the 1930s. We believe the shortfall in SC16 and SC21 is caused by missed active regions owing to data gaps in the \ion{Ca}{2} K observations (see Appendix \ref{app:gaps}), rather than wrongly estimated fluxes in the included plages. This is because for SC21 these have been shown to accurately reproduce the observed magnetograms (Figure \ref{fig:dipoles-nso}).

Comparing Figures \ref{fig:opt-bfly-spots}(a) and \ref{fig:opt-bfly-hale}(a) shows that use of sunspot measurements improves the fit substantially compared to the case where all plage polarities are filled using Hale's Law. In particular, the latter runs seem to underestimate the polar field peak around the mid-1950s (end of SC18), and overestimate the peak around the mid-1960s (end of SC19), hinting that these might be caused by ``rogue'' active regions that don't follow the expected structure. Also, the optimum runs without sunspot measurements show more variation among the different realizations -- this is particularly evident in $b_{1,0}$ (Figure \ref{fig:opt-bfly-hale}b).

\begin{figure*}
    \centering
    \includegraphics[width=\textwidth]{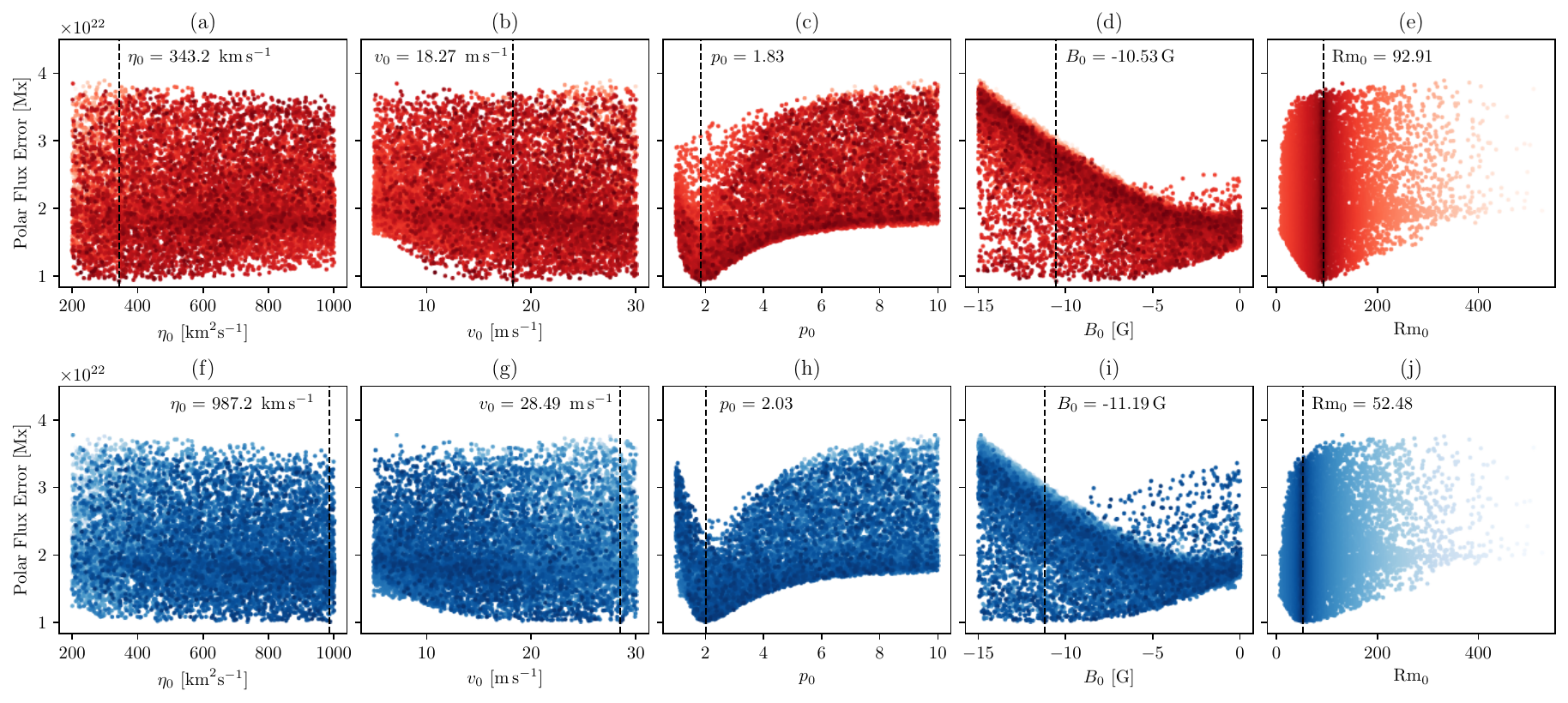}
    \caption{Polar flux error, $E_{\rm PF}$, plotted against the four individual SFT parameters and the dimensionless combination $\mathrm{Rm}_0 = R_\odot^2\Delta_u/\eta_0$, for the single realization of the plage dataset that includes the overall optimum run. Each of the $10\,000$ points shows a single SFT run. The top row shows results using sunspot data, while the bottom row shows results when all polarities are filled using Hale's Law.
    Dashed vertical lines indicate the optimal parameter values (lowest $E_{\rm PF}$) -- these values correspond to Figures \ref{fig:opt-bfly-spots}(c) and \ref{fig:opt-bfly-hale}(c).}
    \label{fig:opt-params}
\end{figure*}

To illustrate how tightly the SFT parameters are constrained, Figure \ref{fig:opt-params} shows $E_{\rm PF}$ for all SFT runs (in a single realization of the plage dataset), projected on each individual SFT parameter in turn. Also shown is the dimensionless combination $\mathrm{Rm}_0$. Optimum parameter values are given by the lowest values of $E_{\rm PF}$, indicated by the dashed vertical lines. This particular realization was selected because it produced the overall lowest value of $E_{\rm PF}$ among the 20 realizations, so these optimum parameter values are the ones used for the runs illustrated in Figures \ref{fig:opt-bfly-spots}(c) and \ref{fig:opt-bfly-hale}(c). The corresponding parameters are shown in the ``Best-fit'' columns of Table \ref{tab:opt-params}, while the ``Plage Ensemble'' columns indicate the variation in these optimum parameters across the different plage realizations.

\begin{deluxetable*}{ccrrrr}
\tablenum{1}
\tablecaption{Optimum SFT parameter values.\label{tab:opt-params}}
\tablewidth{0pt}
\tablehead{
\colhead{Parameter} & \colhead{Units} & \multicolumn2c{(i) With Sunspot Measurements} & \multicolumn2c{(ii) Without Sunspot Measurements}\\
& & \colhead{Best-fit} & \colhead{Plage Ensemble} & \colhead{Best-fit} & \colhead{Plage Ensemble}
}
\startdata
$\eta_0$ & [$\mathrm{km}^2\mathrm{s}^{-1}$] & $343.2$ & $417.1 \pm 198.3$ & $987.2$ & $858.7 \pm 220.4$ \\
$v_0$ & [$\mathrm{m}\,\mathrm{s}^{-1}$] & $18.27$ & $16.70 \pm 2.42$ & $28.49$ & $24.92\pm 4.60$\\
$p_0$ & [none] & $1.83$ & $1.99 \pm 0.14$ & $2.03$ & $3.07\pm 0.69$\\
$B_0$ & [$\mathrm{G}$] & $-10.53$ & $-9.71 \pm 2.44$ & $-11.19$ & $-2.33 \pm 2.74$\\
\hline
$\mathrm{Rm}_0$ & [none] & $92.91$ & $79.91 \pm 18.44$ & $52.48$ & $74.24\pm 50.30$\\
$\lambda_R$ & [$^\circ$] & $5.94$ & $6.58\pm 0.99$ & $7.91$ & $7.34\pm 1.65$\\
\hline
$E_{\rm PF}$ & [$10^{22}\,\mathrm{Mx}$] & $0.93$ & $1.07\pm 0.10$ & $1.01$ & $1.37 \pm 0.17$
\enddata
\end{deluxetable*}

Firstly, observe that the optimum $E_{\rm PF}$ is lower when the plage polarities use sunspot measurements (top row of Figure \ref{fig:opt-params}), compared to the case without (bottom row): $0.93\times 10^{22}\,\mathrm{Mx}$ compared to $1.01\times 10^{22}\,\mathrm{Mx}$. This indicates a better fit and supports the previously mentioned visual impression from Figures \ref{fig:opt-bfly-spots}(a) and \ref{fig:opt-bfly-hale}(a).

Secondly, looking at the top row of Figure \ref{fig:opt-params}, we see that not all of the SFT parameters are equally constrained: the meridional flow shape parameter, $p_0$, is quite strongly constrained around $p_0=1.83$, but a range of values of $\eta_0 \lesssim 600 \,\mathrm{km}^2\mathrm{s}^{-1}$ and $v_0 \gtrsim 12\,\mathrm{m}\,\mathrm{s}^{-1}$ can give comparable values of $E_{\rm PF}$. However, the model is known to depend on the relative strength of flow and diffusion \citep[cf.][]{yeates2023}. Plotting instead the dimensionless ratio $\mathrm{Rm}_0$, as in Figure \ref{fig:opt-params}(e), shows this ratio to be quite tightly constrained around a value of $92.91$. For the run with sunspot data, the initial polar field strength, $B_0$, also has quite a broad acceptable range around $-10.53\,\mathrm{G}$. This suggests that the system is not strongly dependent on this initial condition, which is consistent with the fact that the missing flux in SC16 does not have a lasting detrimental impact throughout the simulation.

Finally, we see from Table \ref{tab:opt-params} that there is some variation in the optimum parameter values across the plage ensemble. For example, $\mathrm{Rm}_0$ has about a $25\%$ standard deviation when sunspot polarity measurements are used, and about a $66\%$ standard deviation when they are not. Generally, we see that the sunspot measurements reduce the uncertainty in all of these optimum parameters, which is encouraging. Notice that there is a relatively smaller uncertainty in $E_{\rm PF}$, the degree of ``optimality'' achieved, compared to that in the parameter values themselves.

%-----------------------------------------------------
\subsection{Validation using polar crown filaments} \label{sec:pcf}

As an independent albeit qualitative validation of our optimum SFT model, the black star symbols in Figure \ref{fig:opt-bfly-spots}(c) show H$\alpha$ observations of the polarmost polar crown filaments, courtesy of \citet{xu2021}. A similar sanity check was shown by \citet{lockwood2022} in their Figure 6. These polar crown filaments tend to form over long-lived East-West neutral lines on the solar surface, and migrate toward the poles at the time of polar field reversal. Because the magnetic field is not perfectly axisymmetric, the locations cannot always be compared directly with $\langle B_r\rangle$. Nevertheless, we see that the timings of the reversals implied by the filament observations, and indeed the implied widths of the polar caps, do compare favorably with the optimum SFT simulation.

%%%%%%%%%%%%%%%%%%%%%%%%%%%%%%%%%%%%%%%%%%%%%%%%%%%%%%%%%%
\section{Evidence for quenching from dynamo effectivity range} \label{sec:effectivity}

Having obtained the best-fit SFT model, we can use it to infer the relative importance of tilt and latitude quenching, thanks to recent work of \citet{talafha2022}. These authors performed a parameter study with a smooth active region source, focusing on the quenching of axial dipole strength, $b_{1,0}$, as a function of cycle amplitude when the source term depends nonlinearly on the cycle amplitude. They compared cases with nonlinearity in the source latitude to those with nonlinearity in the tilts, and measured the deviation of $b_{1,0}$ in each case from a purely linear model. They showed that the relative reduction of $b_{1,0}$ between the two nonlinearities is a function of the ``dynamo effectivity range'',
\begin{equation}
    \lambda_R  = \mathrm{Rm}_0^{-1/2} = \sqrt{\frac{\eta_0}{R_\odot^2\Delta_u}},
    \label{eq:lambdar}
\end{equation}
with latitude quenching dominating if $\lambda_R \lesssim 10^\circ$ but tilt quenching dominating if $\lambda_R$ is larger.

This threshold can be understood from previous work on the contributions of individual active regions within the SFT model \citep{jiang2014,petrovay2020,yeates2023}. Specifically, the amplification of a bipolar magnetic region's axial dipole strength by subsequent SFT evolution is now known to be
\begin{equation}
    \lim_{t\to\infty}\frac{b_{1,0}(t)}{b_{1,0}(t_0)} \approx A\exp\left(-\frac{\lambda_0^2}{2\lambda_R^2}\right),
\end{equation}
where $\lambda_0$ is the region's emergence latitude at $t=t_0$, and $A$ depends on the model parameters. Therefore, only regions emerging within about $\pm \lambda_R$ latitude of the equator will contribute significantly to the end-of-cycle polar field. Latitude quenching dominates in the small $\lambda_R$ regime because regions emerging far from the equator are effectively cut off from contributing to the polar field.

\begin{figure}
    \centering
    \includegraphics[width=\columnwidth]{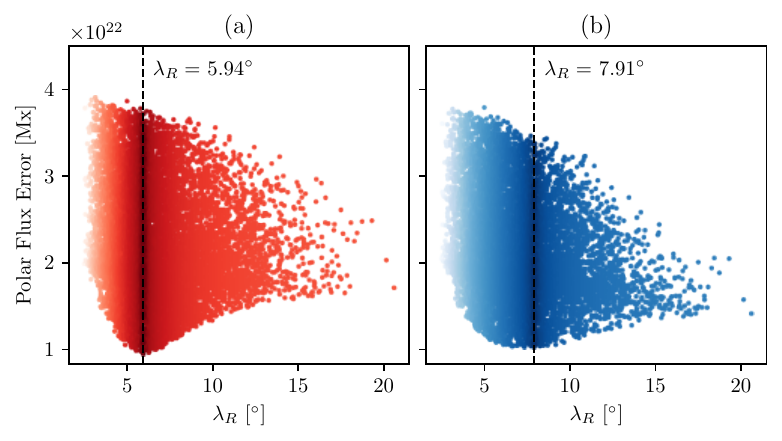}
    \caption{Polar flux error, $E_{\rm PF}$, as a function of the dynamo effectivity range, $\lambda_R = \mathrm{Rm}_0^{-1/2}$, for the single realization of the plage dataset that includes the overall optimum run. As in Figure \ref{fig:opt-params}, each of the $10\,000$ points shows a single SFT run. Panel (a) shows results using sunspot data, while (b) shows results when all polarities are filled using Hale's Law.
    Dashed vertical lines indicate the optimal $\lambda_R$ values.}
    \label{fig:opt-lambda}
\end{figure}

The values of $\lambda_R$ for our best-fit SFT models are shown in Table \ref{tab:opt-params}. (They are given in degrees, obtained by multiplying \eqref{eq:lambdar} by $180/\pi$.) Figure \ref{fig:opt-lambda} shows $E_{\rm PF}$ for the SFT runs projected on this parameter, similar to Figure \ref{fig:opt-params}. When sunspot polarity measurements are used, we find $\lambda_R=5.94^\circ$. The indicated uncertainty from the plage ensemble suggest that the optimum value is safely below $10^\circ$, in the regime where \citet{talafha2022} suggest latitude quenching to be the dominant nonlinearity. Even when sunspot measurements are not used, we see a preference for $\lambda_R\lesssim 10^\circ$, although the precise value is less tightly constrained in that case.

%%%%%%%%%%%%%%%%%%%%%%%%%%%%%%%%%%%%%%%%%%%%%%%%%%%%%%%%%%
\section{Conclusion} \label{sec:conc}

We have presented historical evidence suggesting that -- at least within the Babcock-Leighton model for the solar cycle -- the nonlinear effect of latitude quenching on the solar cycle amplitude is more significant than that of tilt quenching. That is to say, the tendency of active regions to emerge at higher latitudes in stronger cycles has more of a suppressing influence on polar field production than these regions' tendency to be less tilted. 

We have not considered -- and cannot discount -- the possibility of other nonlinearities in the SFT model. For example, flows that change over time in some way dependent on the cycle amplitude. This is not far-fetched; for example, there is evidence for systematic inflows towards active regions \citep{gizon2004}, which could have systematic effects on flux transport and the underlying dynamo \citep[e.g.][]{jiang2010,nagy2020}. But it may be difficult to rule such effects in or out based on the relatively small number of available cycles with polar field measurements. As we have shown, such additional complications -- while not ruled out -- do not appear necessary in order to fit the SFT model to the polar field proxy, provided that realistic enough active regions are used.

Similarly, we have not found it necessary to vary the flux transport parameters from one cycle to the next in order to obtain a reasonable polar field evolution \citep[e.g.][]{wang2002}, nor to add a radial decay term \citep{schrijver2002,baumann2006}. We suggest that the reason these modifications were necessary in previous SFT models could relate to the fact that individual active regions lacked sufficiently detailed observations of their magnetic structure. We have seen the detrimental effect in this paper of throwing away sunspot polarity measurements. And even the assumption of symmetric, idealized bipolar shapes has previously been shown to lead to incorrect estimation of the resulting polar field \citep{iijima2019, jiang2019, yeates2020}.

There are several ways in which we hope to improve or extend our active region database in future. The most obvious would be to extend to the present day (SC25) by cross-calibrating magnetograms from different sources, choosing appropriate thresholds equivalent to our plage regions. We are also hopeful that the database can be extended backward to at least the start of Mount Wilson sunspot measurements in 1917 (CR827), if timing and orientation of the early sunspot measurements can be corrected. For other applications of the SFT model, it would be useful to fill the data gaps, particularly in SC16 -- it may be possible to improve on purely ``synthetic'' filling through (non-trivial) cross-calibration with \ion{Ca}{2} K data from Kodaikanal Solar Observatory \citep{mordvinov2020}, and/or the use of white-light sunspot records.

In summary, we propose that latitude quenching is more important than tilt quenching in regulating the Sun's polar field production. The actual physical mechanism that leads stronger cycles to emerge active regions at higher latitudes is, of course, beyond the scope of our study and requires modelling of the solar interior. One plausible explanation is that flux emergence occurs when the magnetic field strength in the convection zone exceeds some threshold \citep[e.g.][]{cameron2023}. If, further, the poloidal flux is carried gradually equatorward by meridional circulation in the convection zone over the solar cycle as it generates the toroidal flux, then the emergence threshold would be reached at higher latitudes in stronger cycles. Indeed, latitude quenching has been shown to arise within an optimized kinematic ``2x2D'' dynamo model without being explicitly imposed \citep{talafha2022}. But whether this picture is correct remains an open question.

%%%%%%%%%%%%%%%%%%%%%%%%%%%%%%%%%%%%%%%%%%%%%%%%%%%%%%%%%%
\begin{acknowledgments}
This work was funded by UK STFC consortium grant ST/W00108X/1, and benefited greatly from interactions within Teams 420 and 474 at the International Space Science Institute. The National Solar Observatory (NSO) is operated by the Association of Universities for Research in Astronomy (AURA), Inc., under cooperative agreement with the National Science Foundation. We also thank Hisashi Hayakawa, Karen Meyer, and the anonymous reviewer for useful suggestions. MWO calibrated polar faculae data were downloaded from the solar dynamo dataverse (\url{https://dataverse.harvard.edu/dataverse/solardynamo}), maintained by Andrés Muñoz-Jaramillo. The \citet{leussu-data} data were obtained with the  VizieR catalogue access tool, CDS, Strasbourg, France (DOI: 10.26093/cds/vizier). Polar crown filament data were kindly provided by Y. Xu. 
The ensemble datasets of magnetic regions used for the analysis in this paper are published in Durham University's Collection open data repository (\url{http://doi.org/10.15128/r33f462541f}).
\end{acknowledgments}

\software{sunpy \citep{sunpy-code}}

\appendix

%-----------------------------------------------------
\section{Coverage of \ion{Ca}{2} K Observations} \label{app:gaps}

\begin{figure}
    \centering
    \includegraphics[width=0.5\textwidth]{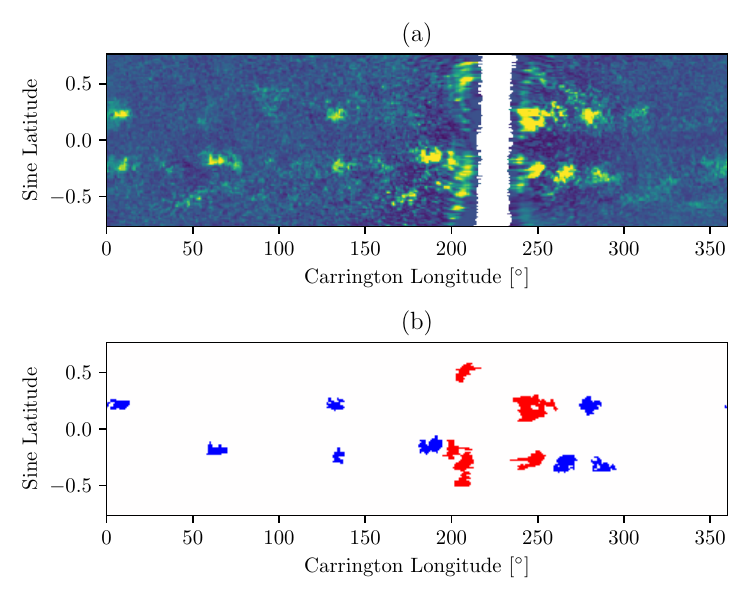}
    \caption{Example of an incomplete \ion{Ca}{2} K synoptic map (a), for CR999. Panel (b) shows the selected plages, with those manually removed shown in red.}
    \label{fig:demo-bad}
\end{figure}

\begin{figure*}
    \centering
    \includegraphics[width=\textwidth]{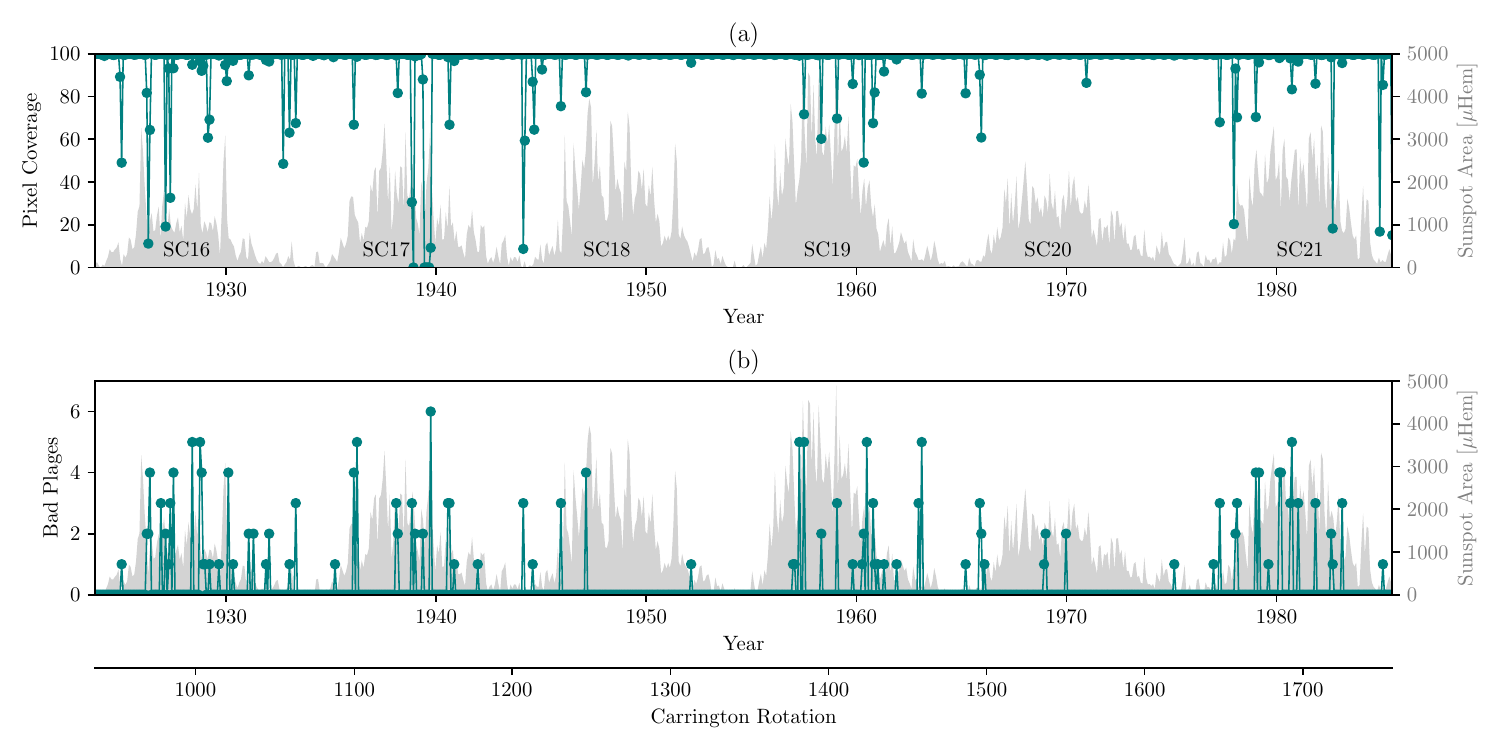}
    \caption{Overview of observational coverage in the Mount Wilson \ion{Ca}{2} K synoptic maps. Panel (a) shows the percentage of pixels in each Carrington map with at least one contributing observation. Panel (b) shows the number of ``bad'' plage regions in each Carrington map, as identified manually. For context, light gray shading shows monthly-averaged sunspot areas from \citet{mandal2020}.}
    \label{fig:cak-gaps}
\end{figure*}

Owing to observational gaps at Mount Wilson Observatory, not all of the \ion{Ca}{2} K synoptic maps in the dataset of \citet{bertello2020} are complete. An example of an incomplete map (CR999, 1928 May to June) is shown in Figure \ref{fig:demo-bad}(a), where only $94\%$ of pixels have at least one contributing observation. (The missing pixels are shown in white.) Figure \ref{fig:cak-gaps}(a) shows this percentage across all of the Carrington rotations. Very few have completely empty synoptic maps, but note that during SC16 and SC21 there are more incomplete maps than at other times. This likely contributes to the shortfall of polar field production in our SFT model during those cycles (Section \ref{sec:opt}).

Figure \ref{fig:cak-gaps}(b) shows the number of ``bad'' plages that were detected in each \ion{Ca}{2} K map. These were manually removed from our subsequent analysis due to being poorly observed. Typically these plages were located on or near the boundary of missing pixel regions in the map. There were 205 in total. The five in CR999 are shown  red in Figure \ref{fig:demo-bad}(b). Note that there were presumably also some (unknown number of) plages located entirely in missing pixel regions, and so completely missed from our dataset. 

%-----------------------------------------------------
\section{Sunspot Longitude Correction} \label{app:offsets}

\begin{figure*}
    \centering
    \includegraphics[width=\textwidth]{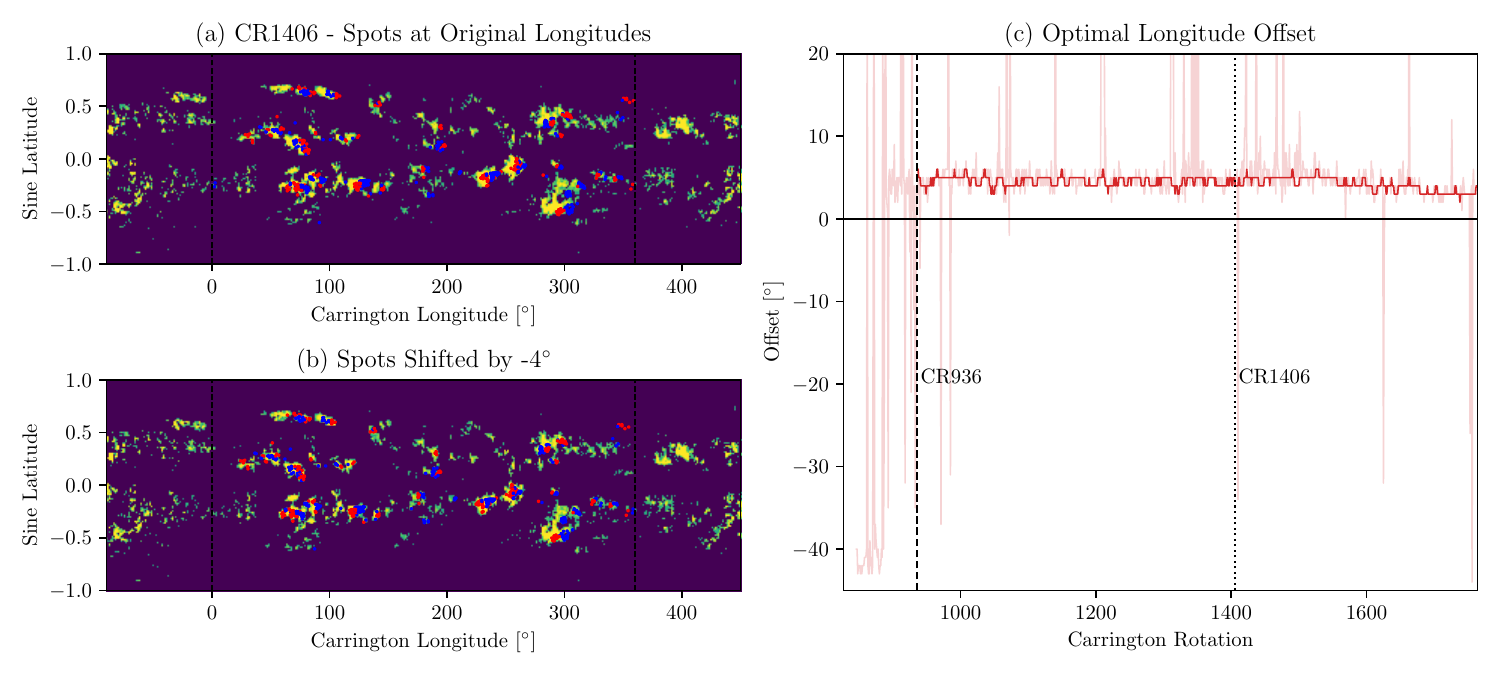}
    \caption{Correction of longitude offsets in the MWO sunspot database. Panel (a) shows the locations of sunspot polarity measurements for CR1406 (red positive, blue negative), overlaid on the \ion{Ca}{2} K synoptic map, while panel (b) shows the same measurements shifted by $-4^\circ$ in longitude. Panel (c) shows the optimal longitude offset for each Carrington rotation -- the faint red curve is the raw estimate by cross-correlating every map, while the darker red curve is the final offset used in the subsequent analysis.}
    \label{fig:spot-offsets}
\end{figure*}

In the digitized sunspot database of \citet{sunspotdata}, each measurement has been assigned a Carrington longitude based on the position relative to central meridian at the recorded time. However, overlaying the spot locations on the \ion{Ca}{2} K synoptic maps shows that these assigned Carrington longitudes appear -- at the time of writing -- to be recorded incorrectly. As a typical example, Figure \ref{fig:spot-offsets}(a) shows the locations of polarity measurements during CR1406 (1958 October to November), overlayed on the \ion{Ca}{2} K synoptic map. By eye the sunspot measurements are offset to the West (right), while Figure \ref{fig:spot-offsets}(b) shows the result of correcting this offset with a uniform $4^\circ$ shift.

The faint red curve in Figure \ref{fig:spot-offsets}(c) shows the optimal longitudinal offset for each rotation in the full dataset (from CR827 to CR1764), determined by cross-correlation between the sunspot locations and the \ion{Ca}{2} K map. We observe a roughly constant offset of around $3^\circ$ to $5^\circ$, superimposed with individual large spikes. After CR936 (1923 September to October), the spikes are due either to incomplete \ion{Ca}{2} K maps, or to rotations with insufficient sunspot measurements to accurately determine the appropriate offset by cross correlation. But before CR936, the sunspot longitudes appear to be unreliable -- with, for example, the same spots recorded at widely varying locations on the Sun on consecutive days. Therefore, for this paper we begin our dataset with CR937.

Since there are also some systematic drifts in the optimum correction over time after CR937 -- for example a reduction since CR1600, we apply a different correction to each Carrington rotation. To account for uncertainty we keep only offsets in the range $[0^\circ, 6^\circ]$ and fill the others by linear interpolation. We then smooth the series with a Savitzky-Golay filter with polynomial degree 1 and a window of width 7 points. (The latter was optimised to give the greatest number of correct polarity pixels in the reconstructed plages for the magnetogram overlap period.) This final offset is shown by the dark red curve in Figure \ref{fig:spot-offsets}.

\bibliography{yeates}{}
\bibliographystyle{aasjournal}

\end{document}